%% file: PART2-testbed_survey.tex
\documentclass[journal]{IEEEtran}
\pdfoutput=1
\IEEEoverridecommandlockouts

\usepackage{acronym}
\usepackage{amsmath,amssymb}
\usepackage[english]{babel}
\usepackage{bbm}
\usepackage{bm}
\usepackage{cite}
\usepackage{enumerate}
\usepackage{graphicx}

\usepackage{mathtools}

\usepackage{lscape}
\usepackage{longtable}
\usepackage{siunitx}
\usepackage{xcolor}
\usepackage{algorithm}
\usepackage{algorithmic}
\usepackage{rotating}

\usepackage[hidelinks]{hyperref}
\usepackage{cleveref}
\usepackage{hhline}

\usepackage{pifont}%
\newcommand{\cmark}{\ding{51}}%
\newcommand{\xmark}{\ding{55}}%

\definecolor{myblue}{RGB}{31,119,180}
\definecolor{myorange}{RGB}{255,127,14}
\definecolor{mygreen}{RGB}{44,160,44}

\acrodef{AHL}{$N$-acyl homoserine lactones}
\acrodef{AI-2}{autoinducer 2}
\acrodef{ATP}{adenosine triphosphate}
\acrodef{BER}{bit error rate}
\acrodef{bioFET}{field-effect transistor biosensor}
\acrodef{C6-HSL}{N-(3-Oxyhexanoyl)-L-homoserine lactone}
\acrodef{CIR}{channel impulse response}
\acrodef{CSK}{concentration shift keying}
\acrodef{DNA}{deoxyribonucleic acid}
\acrodef{FCQD}{fluorescent carbon quantum dot}
\acrodef{EGFET}{electrolyte-gated field-effect transistor}
\acrodef{EM}{electromagnetic waves}
\acrodef{FRET}{F\"orster Resonance Energy Transfer}
\acrodef{FSK}{frequency shift keying}
\acrodef{GFP}{green fluorescent protein}
\acrodef{IoBNT}{Internet of Bio-Nano-Things}
\acrodef{ISI}{inter-symbol interference}
\acrodef{LoC}{lab-on-a-chip}
\acrodef{MC}{Molecular Communications}
\acrodef{MIMO}{multiple-input and multiple-output}
\acrodef{NMC}{Natural Molecular Communication}
\acrodef{OOK}{on-off keying}
\acrodef{RSK}{reaction shift keying}
\acrodef{RNA}{ribonucleic acid}
\acrodef{Rx}{receiver}
\acrodef{sfGFP}{super-folded green fluorescent protein}
\acrodef{SISO}{single-input and single-output}
\acrodef{SMC}{Synthetic Molecular Communication}
\acrodef{SNR}{signal-to-noise ratio}
\acrodef{SPION}{superparamagnetic iron oxide nanoparticle}
\acrodef{ssDNA}{single-stranded DNA}
\acrodef{Tx}{transmitter}
\acrodef{VOC}{volatile organic compound}

\allowdisplaybreaks

\begin{document}
        
\title{Experimental Research\\ in Synthetic Molecular Communications - \\
Part II: Long-Range Communication}

\input{authorblock}

\maketitle

\begin{abstract}
In this second part of our survey on experimental research in \ac{SMC}, we review works on long-range \ac{SMC} systems, i.e., systems with communication ranges of more than a few millimeters.
Despite the importance of experimental research for the evolution of \ac{SMC} towards a mature communication paradigm that will eventually support revolutionary applications beyond the reach of today's prevalent communication paradigms, the existing body of literature is still comparatively sparse.
Long-range \ac{SMC} systems have been proposed in the literature for information transmission in two types of fluid media, liquid and air.
While both types of \ac{SMC} systems, liquid-based and air-based systems, rely on encoding and transmitting information using molecules, they differ substantially in terms of the physical system designs and in the type of applications they are intended for.
In this paper, we present a systematic characterization of experimental works on long-range \ac{SMC} that reveals the major drivers of these works in terms of the respective target applications.
Furthermore, the physical designs for long-range \ac{SMC} proposed in the literature are comprehensively reviewed.
In this way, our survey will contribute to making experimental research in this field more accessible and identifying novel directions for future research.
\end{abstract}
\acresetall

\section{Introduction}

\input{introduction_p2}

\section{Long-Range Liquid-Based SMC}
\label{sec:liquid}
\input{long-range_liquid}

\section{Air-Based SMC}
\label{sec:air}
\input{air}

\section{Conclusions and Outlook}
\label{sec:conclusions}
\input{conclusions_long-range}

\bibliographystyle{IEEEtran}    
\bibliography{IEEEabrv,references}

\end{document}

%% file: authorblock.tex
\author{\IEEEauthorblockN{Sebastian Lotter\IEEEauthorrefmark{1},
    Lukas Brand\IEEEauthorrefmark{1},
    Vahid Jamali\IEEEauthorrefmark{2},
    Maximilian Sch\"afer\IEEEauthorrefmark{1},
    Helene M.~Loos\IEEEauthorrefmark{1}\IEEEauthorrefmark{4},
    Harald Unterweger\IEEEauthorrefmark{3},\\
    Sandra Greiner\IEEEauthorrefmark{1},
    Jens Kirchner\IEEEauthorrefmark{1},
    Christoph Alexiou\IEEEauthorrefmark{3},
    Dietmar Drummer\IEEEauthorrefmark{1},
    Georg Fischer\IEEEauthorrefmark{1},
    Andrea Buettner\IEEEauthorrefmark{1}\IEEEauthorrefmark{4},\\
    and Robert Schober\IEEEauthorrefmark{1}\\} 
    \IEEEauthorblockA{\small\IEEEauthorrefmark{1}Friedrich-Alexander-Universit\"at Erlangen-N\"urnberg (FAU), Erlangen, Germany}\\
    \IEEEauthorblockA{\small\IEEEauthorrefmark{2}Technical University of Darmstadt, Darmstadt, Germany}\\
    \IEEEauthorblockA{\small\IEEEauthorrefmark{4}Fraunhofer Institute for Process Engineering and Packaging IVV, Freising, Germany}\\
    \IEEEauthorblockA{\small\IEEEauthorrefmark{3}Universit\"atsklinikum Erlangen, Section of Experimental Oncology and Nanomedicine (SEON), Erlangen, Germany}
    \vspace*{-1cm}
}

%% file: introduction_p2.tex
In {\em \ac{SMC}}, information is conveyed via the release, propagation, and reception of molecules, called {\em information molecules} in the context of \ac{SMC}.
Specifically and in contrast to conventional communication systems that are based on \ac{EM}, the {\em \ac{Tx}} in \ac{SMC} encodes information into the properties of molecules and the \ac{SMC} {\em \ac{Rx}} senses these properties and decodes the transmitted information.
Molecule properties used for information encoding can be the {\em concentration}, the {\em type}, or the {\em release time} of the molecules.
In the simplest case, binary information is encoded into the molecule concentration by releasing (not releasing) molecules at fixed time instants to transmit a binary '1' ('0'); this modulation scheme is referred to as {\em \ac{OOK}}.
\Ac{SMC} takes place in a fluid medium, liquid or air, and the information molecules can be carried from the \ac{Tx} to the \ac{Rx} by the background flow of the fluid environment, move randomly due to Brownian motion, and possibly undergo chemical reactions.
Also, in some \ac{SMC} systems, information molecules are emitted by the \ac{Tx} in a propulsive manner and exploit their initial velocity for propagation.

\Ac{SMC} has been proposed more than one and a half decades ago as an alternative communication paradigm to enable communication in environments that pose adversarial conditions for \ac{EM}-based communication.
Examples for such environments include the human body, oil pipelines, and industrial facilities that prohibit the use of \ac{EM} due to safety reasons.
Inside the human body, for example, communication among devices of sizes of up to \SI{1}{\micro\meter} may enable novel healthcare applications for early disease diagnosis and targeted drug delivery.
However, the small form factors of the devices and the complex environment of the human body render the implementation of \ac{EM}-based communication difficult.
The fact that natural in-body communication systems such as endocrine signaling rely on molecules for signaling suggests that \ac{SMC} can provide a promising alternative to \ac{EM} for such synthetic in-body applications.
Also, airborne communication via molecules takes place in nature, amongst others in the form of pheromone signaling between insects; air-based \ac{SMC} presents the attempt to adopt this form of communication for synthetic applications.

Since its emergence, \ac{SMC} was explored by researchers using both theoretical and experimental methods.
The main focus of both technical contributions and surveys in the field, however, has been on the theoretical side.
Specifically, the literature on experimental works on \ac{SMC} is comparatively sparse as of today.
Since experimental research will be instrumental for further evolving \ac{SMC} towards a mature and practical communication paradigm, this survey attempts to fill this research gap.
The survey is organized in two parts:

Part~I presents an overview over general concepts in \ac{SMC}, an exposition of the main \ac{SMC} application scenarios, and a comprehensive review of experimental \ac{SMC} works targeting applications with short communication distances, i.e., distances of not more than a few millimeters.

\begin{figure*}
    \centering
    \includegraphics[width=.7\textwidth]{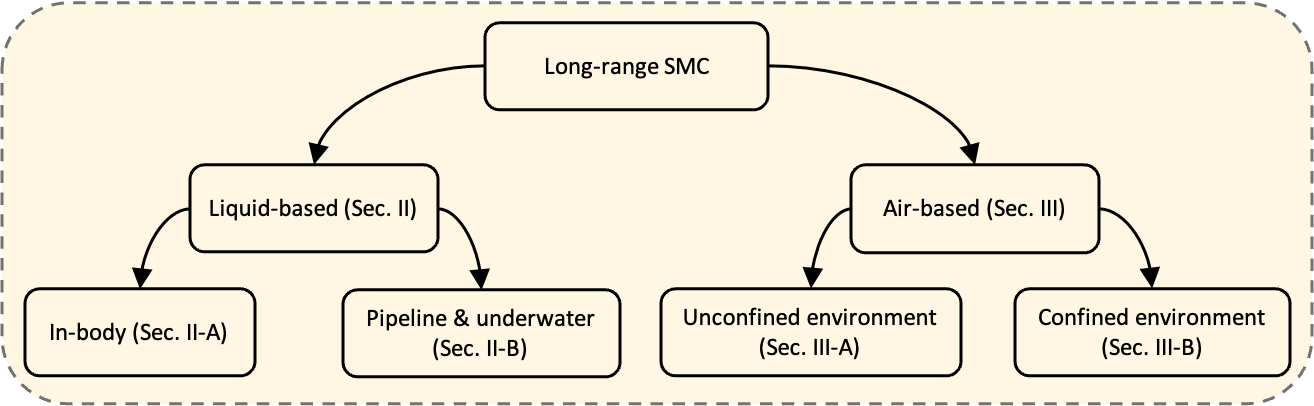}
    \caption{Classification of experimental works on long-range \ac{SMC}.}
    \label{fig:classification}
\end{figure*}

Part II, this paper, presents a comprehensive review of experimental \ac{SMC} works related to {\em long-range} \ac{SMC} applications, i.e., applications requiring communication over distances of more than a few millimeters.
Works in this category are further subdivided based on the type of fluid in which the \ac{SMC} takes place, i.e., {\em liquid} or {\em air}.
Within the class of liquid-based long-range \ac{SMC}, we distinguish between {\em in-body} applications and {\em very large-scale} applications, such as \ac{SMC} in pipelines or the open sea.
Finally, within the class of air-based long-range \ac{SMC}, we distinguish between systems operating within a {\em confined} environment and systems for which the environment is {\em unconfined} (on the scale relevant for \ac{SMC}).
Fig.~\ref{fig:classification} shows the categorization of long-range \ac{SMC} as discussed in this paragraph.
For a complete classification of application scenarios covered in experimental \ac{SMC} works, we refer the reader to Part I of this survey.

The remainder of this paper is organized as follows.
In Sections~\ref{sec:liquid} and \ref{sec:air}, experimental works related to liquid-based and air-based long-range \ac{SMC}, respectively, are reviewed.
Section~\ref{sec:conclusions} concludes this paper with a brief summary of the main results and an outlook towards future research directions.

%% file: long-range_liquid.tex
In this section, we categorize the existing experimental works on long-range liquid-based \acp{SMC} based on whether they are intended for in-body applications or for application scenarios outside the body, e.g., in pipelines or for underwater communication.

\subsection{Long-Range In-Body SMC}

First, we review experimental works in \ac{SMC} that target long-range in-body \ac{SMC} systems.
The inspiration for this type of \ac{SMC} originates from hormone signaling.
Hormone signaling is a \ac{NMC} system that enables communication between distant organs inside the human body via information molecules transported by the blood stream.
On the application side, long-range in-body \ac{SMC} is expected to enable the exchange of information between distant nanosensors deployed inside the human body.
Consequently, the prototypes reviewed in this section mimic the human vascular system by adopting a tube-like physical communication channel.

In contrast to the works on short-range in-body \ac{SMC} reviewed in Part I of this survey, prototypes in the context of long-range in-body \ac{SMC} exclusively build on non-biological \ac{Tx} and \ac{Rx} devices.
In fact, the practical handling of cells or cell colonies is considerably more difficult for the large system dimensions required for long-range \ac{SMC} as compared to short-range \ac{SMC}.
One practical problem here consists of immobilizing functional cells at the high background flow rates typically assumed in long-range \ac{SMC}.
However, the example of hormone signaling shows that long-range in-body communication between biological devices is in principle possible.
So it might be only a matter of time before testbeds for long-range in-body \ac{SMC} with biological \acp{Tx} and/or \acp{Rx} are reported.

\begin{figure*}
    \centering
    \includegraphics[width=.8\textwidth]{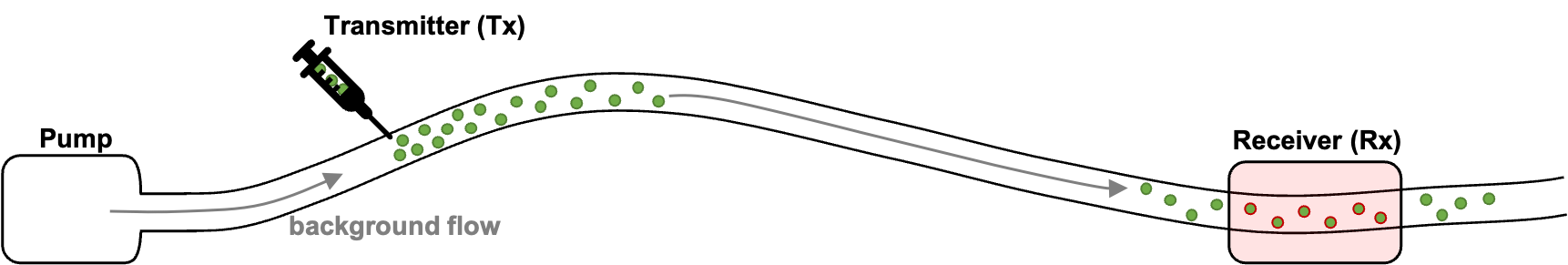}
    \caption{Prototypical setup for long-range in-body \ac{SMC}. A background flow in a tube-shaped channel is generated by a pump. The \ac{Tx} injects information molecules (green) that are sensed by the \ac{Rx}. Sensed molecules are depicted by green disks with red boundaries.}
    \label{fig:long-range_in-body}
\end{figure*}

In the rest of this section, we first review the main experimental end-to-end prototypes for this class of systems along with their various extensions.
Then, we discuss two recent experimental works that focus specifically on the \ac{Tx} and \ac{Rx} design, respectively.

\subsubsection{End-To-End Prototypes}

The end-to-end testbeds for long-range in-body \ac{SMC} proposed in the literature share the common pattern that in all of them information molecules are released by injection via a syringe or a valve into a flowing liquid contained in a tube structure.
The liquid carries the information molecules towards an \ac{Rx} at which the output signal is generated.
For a schematic depiction of this setup, see Fig.~\ref{fig:long-range_in-body}.
We could identify six main testbed architectures in the literature \cite{farsad17,tuccitto17,unterweger18,koo20,wang20a,pan22} from which further variations have evolved.

The six architectures discussed in the following mainly differ in the choice of information molecules which, in turn, crucially influences the choice of the physical \ac{Rx}.
Table~\ref{tab:lr_inbody:sigmol_rx} shows the different information molecule/\ac{Rx} combinations employed in these prototypes.
Furthermore, it is indicated in Table~\ref{tab:lr_inbody:sigmol_rx} whether the employed information molecules are soluble in the medium and whether the \ac{SMC} signal is detected invasively or non-invasively by the \ac{Rx}, cf.~Section II in Part I of this survey.

\begin{table*}
    \centering
    \caption{Information molecule/physical Rx combinations in experimental prototypes for long-range in-body \ac{SMC}.}
    \begin{tabular}[t]{|c|c|c|c|c|}
        \hline
        \textbf{Ref. (Year)} & \textbf{Information molecules} & \textbf{Information molecules soluble?} & \textbf{Physical Rx} & \textbf{Invasive Rx?} \\\hline
        \cite{farsad17} (2017) & Hydrogen ions & \cmark & pH probe & \cmark \\\hline
        \cite{tuccitto17} (2017) & Fluorescent nanoparticles & \xmark & LED + photodiode & \xmark \\\hline
        \cite{unterweger18} (2018) & Superparamagnetic iron oxide
        nanoparticles & \xmark & Susceptometer & \xmark \\\hline
        \cite{koo20} (2020) & Glucose & \cmark & Electrolyte-gated FET & \cmark \\\hline 
        \cite{wang20a} (2020) & Tablesalt & \cmark & Electrical conductivity reader & \cmark \\\hline
        \cite{pan22} (2022) & Color pigments & \cmark & Color sensor & \xmark \\\hline
    \end{tabular}
    \label{tab:lr_inbody:sigmol_rx}
\end{table*}

\textbf{pH-based end-to-end \ac{SMC}:}
In \cite{farsad17}, {\em acidic} and {\em basic} solutions were used to encode binary data.
An acidic solution is an aqueous solution with a pH value less than 7, while the term basic solution refers to a solution with pH value larger than 7.
Specifically, a binary '1' was transmitted in \cite{farsad17} by pumping a pulse of basic liquid into the background flow, where the background flow consisted of water with a pH value ranging from $7$ to $8.5$.
A binary '0', in turn, was transmitted by injecting a pulse of acidic solution.
The resulting change in the pH value of the medium, i.e., the received signal, was recorded using a pH meter, a device that measures the electro-chemical potential difference between a reference solution and the solution being probed.
Several different detectors were proposed in \cite{farsad17} to recover the binary data from the received signal.
It was found that machine learning-based detectors outperform the baseline detection scheme based on the slope of the received signal by a factor of $10$ in terms of the achieved \acp{BER}.
The comparatively good performance of the machine learning-based detectors in \cite{farsad17} was attributed to the capability of these detectors to mitigate the interference between subsequent injections of information molecules.
Model-based detectors were not considered in \cite{farsad17}, since the chemical interactions of molecules in the medium and the unknown sensor noise at the \ac{Rx} rendered the derivation of analytical models for the proposed system difficult.

The testbed proposed in \cite{farsad17} was later used to evaluate the performance of machine learning-based detection methods for \ac{SMC} proposed in \cite{farsad18}.
Furthermore, a \ac{MIMO} \ac{SMC} system was proposed in \cite{lee20} based on the experimental platform discussed in \cite{farsad17}.
In particular, four peristaltic pumps were used in \cite{lee20} to transmit two data streams concurrently using binary time shift-keying.
Here, each symbol interval is divided into two time slots and an acidic solution is injected into the channel in either of the two time slots depending on whether a binary '1' or '0' is transmitted.
This type of modulation is called {\em pulse position modulation}.
Furthermore, the physical channel from \cite{farsad17} is extended by introducing several junctions in \cite{lee20} and two pH meters are employed by the \ac{Rx} in \cite{lee20} instead of a single one in \cite{farsad17}.
However, a comprehensive performance analysis of the testbed proposed in \cite{lee20} is still missing.

\textbf{Fluorescence-based end-to-end \ac{SMC}:}
In the long-range \ac{SMC} testbed presented in \cite{tuccitto17}, {\em \acp{FCQD}} were proposed as information molecules.
\acp{FCQD} are a class of fluorescent, biocompatible nanoparticles that can, for example, be synthesized from citric acid.
\Acp{FCQD} exhibit fluorescence upon irradiation with light of a particular wavelength, which is absorbed in the process (absorbance).
Moreover, \acp{FCQD} show {\em self-quenching}, i.e., reduced light emission, when they aggregate in large concentrations.
This non-linear dependence of the fluorescence emission on the concentration of \acp{FCQD} was exploited in the fluorescence-based system proposed in \cite{tuccitto17} to enhance the communication range as compared to absorbance-based signaling.

The testbed proposed in \cite{tuccitto17} was extended in \cite{tuccitto18} to exploit pH-dependent chemical reactions of information molecules in the medium.
To this end, fluorescein diacetate was adopted as information molecule in \cite{tuccitto18}.
Fluorescein diacetate submersed in a pH buffer solution undergoes hydrolysis in a pH-dependent manner and converts to fluorescein, a fluorescent molecule.
By tuning the pH value of the medium, the received signal could be optimized for different distances between \ac{Tx} and \ac{Rx}.

The system designs from \cite{tuccitto17} and \cite{tuccitto18} were combined in \cite{fichera19} to implement long-range \ac{SMC} based on \ac{RSK}.
To this end, two types of information molecules were employed, \acp{FCQD} and copper ions.
In the presence of copper ions, \acp{FCQD} are quenched, i.e., their fluorescence emission is decreased.
The absorbance spectrum of the \acp{FCQD}, however, is unaltered by the presence of the copper ions.
Exploiting this effect, \acp{FCQD} are used for synchronization of the \ac{Tx} and the \ac{Rx} by means of the absorbance signal in \cite{fichera19}, while the copper ions are injected to modulate binary information onto the fluorescence signal.
Consequently, the fluorescence signal is used for detection in \cite{fichera19}.
The testbed first introduced in \cite{tuccitto17} was further extended in \cite{lidestri19,fichera20,fichera21,cali21,cali22a,cali22b}, and a comprehensive description of the experimental methodology used in these works was published in \cite{cali22}.

\textbf{Magnetic susceptibility-based end-to-end \ac{SMC}:}
The long-range \ac{SMC} testbed proposed in \cite{unterweger18} is based on the use of {\em \acp{SPION}} as information molecules.
At the \ac{Rx}, the magnetic property of the \acp{SPION} is exploited for signal detection with a susceptometer, i.e., a coil surrounding the transmission tube at the \ac{Rx} site.
Namely, as the \acp{SPION} pass through the susceptometer, the susceptometer detects the change in the magnetic susceptibility of the transmission medium induced by the presence of the \acp{SPION}.
Hence, similar to the testbed design in \cite{tuccitto17} and in contrast to \cite{farsad17}, the received signal is measured non-invasively in \cite{unterweger18}, cf.~Section II in Part I of this survey.

The \ac{SPION}-based testbed originally introduced in \cite{unterweger18} was further developed and comprehensively studied in \cite{bartunik19,ahmed19,bartunik20b,bartunik21,wicke22}.
An improved implementation for the \ac{Rx} originally presented in \cite{unterweger18} was introduced in \cite{bartunik19}.
The physical \ac{Rx} employed in \cite{unterweger18} was characterized in \cite{ahmed19} with respect to the impact of the position of the \acp{SPION} inside the susceptometer on the received signal.
In \cite{bartunik20b}, the \ac{OOK} modulation scheme used in \cite{unterweger18} was extended to $6$-ary \ac{CSK}.
Using machine learning-based signal detection, the data rate in \cite{bartunik20b} was almost doubled as compared to \cite{unterweger18}, at a \ac{BER} of 1\%.
In \cite{bartunik21}, the testbed from \cite{unterweger18} was extended to a \ac{MIMO} system by introducing fluorescent dyes as an additional type of information molecules besides \acp{SPION}.
Consequently, unlike the \ac{Rx} in \cite{unterweger18}, the \ac{Rx} in \cite{bartunik21} comprises not only a susceptometer, but also a fluorescence detector to measure the concentration of fluorescent dyes.
It was shown in \cite{bartunik21} that nearly orthogonal transmission could be achieved between the \ac{SPION}-based and the fluorescent dye-based channels, leading to an increase in the achievable data rate as compared to \cite{unterweger18}.
Finally, the testbed introduced in \cite{unterweger18} was analyzed comprehensively from a communication theoretic perspective in \cite{wicke22}.
The results of this analysis were exploited to devise novel detection methods which, in turn, were validated with the experimental testbed data.

\textbf{Glucose-based end-to-end \ac{SMC}:}
In the testbed presented in \cite{koo20}, glucose molecules were employed as information molecules and an \ac{EGFET} was proposed as physical \ac{Rx} for in-body \ac{SMC} between implantable biosensors.
Binary data was modulated onto the injection of glucose molecules into the medium using \ac{OOK}.
The glucose concentration was measured at the \ac{Rx} site by a transistor coated with glucose oxidase molecules, a so-called \ac{EGFET}.
Hereby, the received signal generated at the \ac{EGFET} results from a local change in the pH value caused by the oxidation of glucose that is catalyzed by the immobilized glucose oxidase molecules.
In order to measure the glucose molecule concentration, the physical \ac{Rx} in \cite{koo20} was immersed into the medium.
The authors employed different machine learning-based detection methods and compared the system performances in terms of the \ac{BER} for different data rates.
Interestingly, they found that for the proposed {\em module-based universal detector} consisting of several different machine learning-based modules, an increased data rate can lead to a reduced \ac{BER}.
This phenomenon is attributed in \cite{koo20} to the positive impact of increased liquid injection on the suppression of interference between subsequent transmissions at the \ac{Rx} via the increased washing-out of residual information molecules from the reactive \ac{EGFET} surface.

\textbf{Electrical conductivity-based end-to-end \ac{SMC}:}
In \cite{wang20a}, a long-range in-body \ac{SMC} testbed was introduced that employs tablesalt (NaCl) as information molecule.
Binary data is modulated onto the injection of a saline solution into the aqueous medium using \ac{OOK}.
The signal is detected in \cite{wang20a} by an electrical conductivity reader immersed into the physical communication channel at the \ac{Rx} site.
Furthermore, in the testbed proposed in \cite{wang20a}, the \ac{Tx} and the \ac{Rx} are connected via two tubes of different lengths.

The authors in \cite{wang20a} proposed a novel joint detection and estimation scheme to address the issues of non-causal \ac{ISI} and long delay spread relative to the channel coherence time.
The proposed scheme is based on an adaptive Viterbi decoder and it is applied to detect binary data transmitted using the testbed setup described above.
The authors show that with the proposed scheme a \ac{BER} of $0.1$ can be achieved for the proposed two-path testbed at a data rate of 10 bits per second.

\textbf{Color-based end-to-end \ac{SMC}:}
Finally, in \cite{pan22}, color pigments were employed as information carrying molecules and detected non-invasively by a color sensor at the \ac{Rx} site.
In contrast to non-soluble information molecules, for example \acp{SPION} \cite{unterweger18}, the color pigments dissolve in water and are seamlessly carried by the background flow.
Hence, sedimentation of information molecules on the bottom of the channel was minimal in \cite{pan22}.
The feasibility of the proposed setup to achieve reliable transmission of digital data using \ac{OOK} at a data rate of $0.1\,\mathrm{bps}$ (bit per second) was demonstrated in \cite{pan22}.
Furthermore, the pressure and the flow rate of the medium were measured non-invasively, further characterizing the proposed system.

While all experimental works reviewed so far in this section rely on more or less bulky devices for signal transmission and reception, which may be difficult to use in practical in-body applications, the works reviewed in the following section focus on the design of \ac{Tx} and \ac{Rx} devices that overcome this issue.

\subsubsection{Tx and Rx Design} In the following, we discuss a \ac{Tx} and an \ac{Rx} implementation. 

\textbf{ssDNA-coated electrode-based Tx:} In \cite{luo20}, a \ac{Tx} for long-range in-body \ac{SMC} in the form of an electrode coated with a multilayer film  was proposed where each layer of the film in turn is coated with \ac{ssDNA}.
In the proposed \ac{Tx} design, the molecular signal in the form of \ac{ssDNA} molecules released from the coated electrode into the surrounding liquid medium is generated in response to an electrical signal applied to the electrode.
Specifically, it was demonstrated in \cite{luo20} that the applied electrical signal causes a gradual release of \ac{ssDNA} from the \ac{Tx}.
Furthermore, by repeatedly sampling the bulk \ac{ssDNA} concentration in the medium it was shown in \cite{luo20} that a stream of digital data could be transmitted with the proposed \ac{Tx} design using \ac{OOK}.

\textbf{Graphene-based bioFET-based Rx:} A graphene-based \ac{bioFET} functionalized with \ac{ssDNA} as \ac{SMC} \ac{Rx} was developed and analyzed in \cite{kuscu21}.
In the experimental setup in \cite{kuscu21}, the proposed \ac{bioFET} was embedded into a microfluidic chamber into which information molecules in the form of \ac{ssDNA} matching the \ac{ssDNA} used for functionalization of the \ac{bioFET} could be introduced in a controlled manner.
It was shown that the proposed \ac{bioFET} is responsive to the presence of information molecules in terms of its source-drain current.
Furthermore, it was demonstrated in \cite{kuscu21} that binary data modulated onto the concentration of information molecules using \ac{OOK} could be successfully detected from the received signal at the \ac{bioFET}.
The particularly small form factor and the two-dimensional geometry of the proposed \ac{Rx} design make it a compelling candidate for the implementation of nanodevices for long-range in-body \ac{SMC}.

\subsection{Pipeline and Underwater Communications}

In this section, we review experimental works on very large-scale liquid-based \ac{SMC}.

\textbf{Flow-aided SMC in pipeline structures:} References \cite{atthanayake18,khaloopour19a,abbaszadeh19} focus on flow-aided transmission in pipeline-like duct geometries.

In \cite{atthanayake18}, the authors studied the impact of obstacles of different shapes on the signal transmission in a rectangular duct.
To this end, fluorescent dyes were injected into a water tank to transmit binary data using \ac{OOK}.
The concentration of fluorescent dyes at the \ac{Rx} site, \SI{5}{\meter} downstream of the injection site, was measured using two different methods.
One set of concentration measurements was obtained from camera footage using particle image velocimetry.
The other set of measurements was recorded with a fluorometer.
Both measurement devices, the camera and the fluorometer, were placed inside the water tank.
The water in the tank was subject to flow with different flow rates directed from the \ac{Tx} towards the \ac{Rx}.

Exploiting the measurement data obtained in the setup described above, the authors in \cite{atthanayake18} proposed an additive Gaussian noise model for the received signal and conducted a performance analysis with respect to the received \ac{SNR} and the Shannon capacity of the system.
Evaluating the impact of different types of obstacles on the system performance as compared to the case without obstacles, the authors found that the system performance did not deteriorate in the presence of obstacles.
For some cases, even a positive impact of obstacles on the system performance was observed.
The authors in \cite{atthanayake18} argued that this positive effect resulted from the emergence of stable structures, i.e., vortex rings, in the presence of the turbulent flow introduced by the obstacle.
Similar phenomena have been observed in air-based \ac{SMC} experiments in which the transmission medium exhibits turbulent flow \cite{ozmen18,abbaszadeh18}, cf.~Section~\ref{sec:air}.

In \cite{abbaszadeh19}, the noise distribution of the received signal in an experimental setup very similar to the one in \cite{atthanayake18} was studied.
However, in \cite{abbaszadeh19} the statistics of the fluorescence signal emitted by the fluorescent dyes in response to illumination by a planar laser source was studied.
In particular, novel noise models under steady and turbulent flow conditions, respectively, were proposed in \cite{abbaszadeh19} based on the experimental observations.

In \cite{khaloopour19a}, a large-scale liquid-based \ac{SMC} testbed was proposed.
In contrast to \cite{atthanayake18,abbaszadeh19}, the testbed in \cite{khaloopour19a} focuses on the laminar flow regime.
In \cite{khaloopour19a}, binary information is encoded into the injection of an acidic solution into a pipe with radius \SI{10}{\centi\meter} and length \SI{2}{\meter} using \ac{OOK}.
The background flow in the pipe consists of water with a pH value of 7.
The change in pH value resulting from the injection of the acidic solution into this background flow is measured using a pH meter.

In addition to experimental investigation, comprehensive modeling was performed in \cite{khaloopour19a}.
This included analytical and simulation models to determine the \ac{CIR} of the proposed \ac{SMC} system, derive the detector, and compute the achievable \ac{BER} at a given transmission rate.
The results obtained from the proposed analytical and simulation models were compared with experimental data and good agreement was observed.
For the proposed \ac{SMC} setup, \acp{BER} of approximately $2$ and $6$ percent for transmission rates of $1/27\,\si{\per\second}$ and $1/22\,\si{\per\second}$, respectively, could be achieved in \cite{khaloopour19a}.

\textbf{Underwater SMC without flow:} Some experimental works on large-scale \ac{SMC} have considered signal transmission for underwater communication where the signal propagation does not rely on background flow that is directed towards the \ac{Rx} \cite{huang19,guo20}.

In \cite{huang19}, a \ac{MIMO} \ac{SMC} system is proposed, where an acidic solution is injected by one out of four \ac{Tx} tubes into a small-sized box filled with water.
The information in this setup is modulated onto the index of the \ac{Tx} tube by which the solution is injected.
At the \ac{Rx} site, four pH probes are positioned opposite to the \ac{Tx} tubes to detect the change in pH value in response to the injection of the acidic solution.
The signal propagation in \cite{huang19} relies on the diffusion of protons in the medium and on their initial velocity after injection.

In \cite{guo20}, the physical phenomena of buoyancy and gravity are explored for underwater \ac{SMC}.
In particular, an experimental setup was presented in \cite{guo20} to explore the possibility of vertical underwater \ac{SMC}.
In this setup, silver-coated hollow glass spheres were injected as information carrying information molecules into a water tank.
The particles propagated by buoyancy and gravity, and were carried by the background flow in the tank.
Furthermore, they were tracked using a camera.
The results presented in \cite{guo20} suggest a statistical model for the velocity distribution of the information molecules following a Gaussian distribution.

\subsection{Summary}

In this section, we first reviewed different experimental works on long-range in-body \ac{SMC}.
We identified six basic types of long-range in-body \ac{SMC} testbeds and discussed several extensions to these main testbeds.
The system designs of these six main testbeds differ mainly in the choice of information molecules and physical \acp{Rx}.
Some physical \acp{Rx} proposed in the literature require invasive measurement of the received \ac{SMC} signal, while others do not.
Some information molecules dissolve in the carrier liquid, others do not, cf.~Table~\ref{tab:lr_inbody:sigmol_rx}.
Depending on the specific application, some of these properties may be desirable in practice, and hence guide the system design.
We also discussed two works that focus on the design of a practical \ac{Tx} and a practical \ac{Rx} for long-range in-body \ac{SMC}.
Finally, we discussed some experimental works on pipeline and underwater \ac{SMC}.
While first promising results were achieved in this context, the suitability of the proposed \ac{SMC} schemes for practical applications remains to be shown.

%% file: air.tex
\begin{figure*}
    \centering
    \includegraphics[width=.7\textwidth]{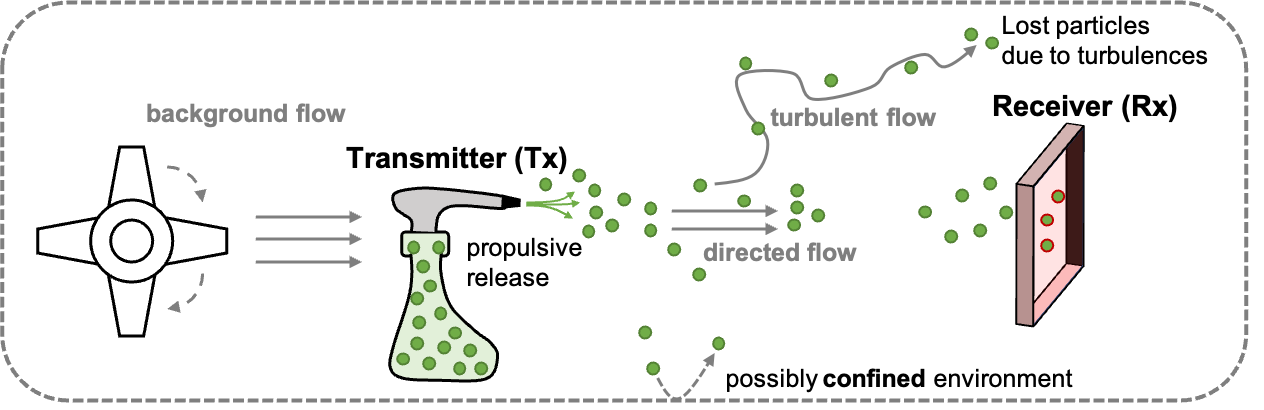}
    \caption{Prototypical setup for air-based \ac{SMC}. Background flow realized by a fan supports the molecule propagation. Some of the experiments are implemented in confined environments, which additionally impact the molecule propagation. There exist different mechanisms for releasing the information molecules, cf. Table~\ref{tab:air_based:release_mechanism} and Fig.~\ref{fig:air:release_mech}. After propagation, the molecules may be detected by the \ac{Rx}.}
    \label{fig:air_based_MC}
\end{figure*}

In this section, we review experimental studies of air-based \ac{SMC}.
We first give an outline of applications that motivate the development of these \ac{SMC} systems.
Then, we review and discuss experiments that explore air-based \ac{SMC} in unbounded and bounded domains, respectively, cf.~Fig.~\ref{fig:air_based_MC}. Hence, in this survey, we classify air-based \ac{SMC} applications upon the existence of boundaries in their environment, which we denote as \textit{confined environment} and \textit{unconfined environment}, respectively.
Next to the environment, for experimental air-based \ac{SMC}, different release mechanisms and sensing principles have been reported, which we summarize in Table~\ref{tab:air_based:release_mechanism}.

\subsection{Main Application Scenarios}

\begin{table*}
    \centering
    \caption{Release mechanism / physical Rx combinations in experimental prototypes for air-based \ac{SMC}.}
    \begin{tabular}[t]{|c|c|c|c|c|c|}
        \hline
        Release mechanism vs. sensing principle  & \parbox{1.5cm}{\textbf{Metal \\ oxidation}} & \parbox{1.8cm}{\textbf{Mass \\ spectrometer}} & \parbox{1.8cm}{\textbf{Photo \\ionization}} & \textbf{Fluorescence} & \parbox{2.5cm}{\textbf{Surface acoustic \\ waves}} \\\hline
        \textbf{Spray (nebulized information molecules)} & \cmark \cite{farsad13, koo16, qiu16, qiu14} & \xmark & \xmark & \cmark \cite{schurwanz21, bhattacharjee20} & \xmark \\\hline
        \textbf{Air stream (vaporized information molecules)} & \cmark\cite{purnamadjaja07} & \cmark \cite{mcguiness18, mcguiness19, giannoukos17} & \cmark \cite{ozmen18, shakya18} & \xmark & \cmark \cite{cole09} \\\hline
        \textbf{Fluid droplet (dissolved information molecules)} & \cmark \cite{sousa2008toward} & \xmark & \xmark &  \xmark & \xmark \\\hline
    \end{tabular}
    \label{tab:air_based:release_mechanism}
\end{table*}

\begin{figure}
    \centering
    \includegraphics[width=0.9\linewidth]{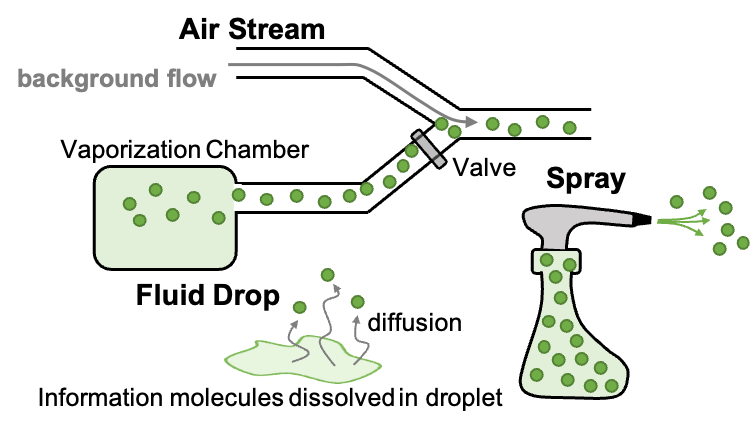}
    \caption{The three different release mechanism used in experimental work on air-based \ac{SMC}. Molecules are either released by a spray, in vaporized form via an air stream, or dissolved via a fluid drop, cf. Table~\ref{tab:air_based:release_mechanism}.}
    \label{fig:air:release_mech}
\end{figure}

The inspiration for many air-based \ac{SMC} applications is drawn from airborne \ac{NMC} systems; examples include pheromone communication and odor communication, where chemo-communicative substances, pheromones and odors, respectively, are used as information molecules.
In pheromone communication, the behavior of individuals is influenced, which is especially well studied for insects.
Different types of pheromones exist for different purposes like communication of threat, food sources, and reproductive states. 
Understanding the release, propagation, and reception of airborne pheromones and other volatile communication substances helps to develop future air-based \ac{SMC} systems for communication with living organisms. In the long run, studying air-based \ac{NMC} systems will provide the necessary chemical, ecological, and technological background to ensure safe use and high effectiveness of \ac{SMC} for mating disruption or mass trapping of insects in agriculture.

Moreover, studying air-based \ac{SMC} enables the development of nature-inspired and model-driven \ac{Tx} and \ac{Rx} systems, in both pure \ac{SMC} systems but also in interfaces between living organisms and technical systems.
For example, technical systems for odor detection, like electronic noses, have been developed and are constantly improved.
In future, such systems could allow patients suffering from anosmia, i.e., smell blindness, to smell.

An additional motivation of air-based \ac{SMC} in an unconfined environment is to enable long-range wireless communication in emergency situations, where electromagnetic and optical communication is not available.
In particular, similar to \ac{NMC} based on chemo-communicative substances, the operation of a robot swarm in a search and rescue operation is envisioned to be coordinated by \ac{SMC}, where one robot upon reaching a missing person guides the other robots by releasing information molecules \cite{purnamadjaja07}.
In contrast to EM, information molecules do not disappear immediately and remain in fractions in the environment, which can be exploited for pathfinding, i.e., synthetic chemo-communicative substances may help to mark already explored areas and dangerous paths \cite{sousa2008toward}.

In confined environments, e.g., in tunnels, caves, and (ventilation-)pipe networks, EM-based communication suffers from high path losses; air-based \ac{SMC} systems may help to overcome this issue \cite{qiu14}.

\subsection{Experimental Works}

In this section, experiments for air-based \ac{SMC} are discussed where information was transmitted end-to-end over a physical unbounded (Section~\ref{air:unconfined}), and bounded (Section~\ref{air:bounded}) channel, respectively.
Note that also indoor experiments are denoted as unconfined, if the boundaries are far away relatively to the scale of the experimental setup, and therefore do not impair the results of the experiments.

\subsubsection{Unconfined Environments}\label{air:unconfined}

In unconfined air-based \ac{SMC}, cf. Fig.~\ref{fig:air_based_MC}, the information molecules are released in liquid or gaseous forms by a \ac{Tx} via spraying \cite{farsad13, koo16, schurwanz21,qiu16}, a liquid drop \cite{sousa2008toward}, or a stream of air carrying the chemical vapor \cite{mcguiness18, purnamadjaja07}. The transmission of the information molecules is possibly aided by background airflow towards the \ac{Rx}, e.g., by a fan \cite{farsad13, purnamadjaja07, koo16}, or by an initial force at the \ac{Tx}. This initial force results from the impulsive force in spraying or the large flow speed of the air stream.

\textbf{Mobile air stream-based transmission:} One of the earliest experiments, which can be characterized as air-based \ac{SMC}, is \cite{purnamadjaja07}, where, inspired by natural pheromone communication, synthetic pheromone-based communication was studied in the context of autonomous robots.
In particular, in a leader-follower scenario a group of mobile robots was guided by a leading robot, which triggers different cooperative behaviors encoded in different chemicals, namely methylated spirits and eucalyptus oil.
The leading robot was equipped with a pump, containers, and fan for steering the released molecules.
Furthermore, all robots carried a metal oxide gas sensor for detecting the
chemicals and an airflow sensor to measure wind velocity and its direction.

\textbf{Mobile fluid droplet-based transmission:} In \cite{sousa2008toward}, volatile information molecules dissolved in ethanol were placed as droplets on the ground for path marking and finding. Mobile robots equipped with an artificial olfactory system searched for these volatiles, which vaporize over time. The authors show that the detection reliability can be improved by a fan oriented to the surface, near the \ac{Rx} unit of the mobile robots, for which a metal oxide gas sensor was used.

\textbf{Spray-based transmission:} In contrast to \cite{purnamadjaja07, sousa2008toward}, where chemicals were used for coordinating behavior and localization purposes, in \cite{farsad13} air-based \ac{SMC} was used for an end-to-end transmission of text messages.
The text was converted to binary data, which modulate the emission of isopropyl alcohol from an electronically controlled mechanical sprayer.
The signal was detected by a metal oxide sensor at the \ac{Rx} site, which was placed up to \SI{4}{\meter} away from the \ac{Tx}.

The information transmission was evaluated in the presence and in the absence of flow directed from the \ac{Tx} towards the \ac{Rx} and induced by a fan.
As the testbed in \cite{farsad13} is easily accessible, consisting of off-the-shelf components, many extensions have been proposed since its first appearance.
In the following, we highlight some of the extensions to \cite{farsad13}.

\textbf{Multiple sprays-based transmission:} To achieve higher data rates compared to \cite{farsad13}, \ac{MIMO} air-based \ac{SMC} systems were proposed and validated for the first time experimentally in \cite{koo16}. In particular in \cite{koo16}, a 2×2 molecular \ac{MIMO} system was implemented using algorithms, which were obtained via a \ac{MIMO} simulator prior to this. The authors showed that data rates of $0.34\,\mathrm{bps}$ at an error level of $0.0975$ can be achieved for a system with a distance of \SI{90}{\centi\meter} between opposite \acp{Tx} and \acp{Rx} and \SI{40}{\centi\meter} between adjacent \acp{Tx} and \acp{Rx}, respectively.

\textbf{Air stream-based transmission:} In \cite{mcguiness18}, a different realization for the \ac{Tx} and \ac{Rx} was proposed. The authors experimented with an evaporation chamber as \ac{Tx}, which releases \acp{VOC} as information molecules, and a membrane inlet mass spectrometer as \ac{Rx}, which allows for the detection and identification of a high number of different molecules. In a mass spectrometer, the molecules are ionized, separated based on their mass-to-charge fractions, and detected. The results are generally represented visually as mass spectrum. The system has also been studied by the same working group for a confined environment \cite{mcguiness19,giannoukos17}.
 
\textbf{Mobile spray-based transmission:} In \cite{qiu16}, the tabletop experiment in \cite{farsad13} was extended to the case where the \ac{Tx} is mobile. The authors showed experimentally that for scenarios, where the molecule diffusion was slower than the movement of the \ac{Tx}, the \ac{Rx} may detect consecutive bits in an interchanged order. To cope with that challenge, positional distance channel codes were proposed. Computer-based simulations showed a better performance in terms of \ac{BER} of these codes compared to classical Hamming-distance codes.

\textbf{Fluorescent spray-based transmission with high initial velocity:} Different to all previously discussed air-based \ac{SMC} systems, in \cite{schurwanz21}, experimental research driven by air-based \ac{SMC} was conducted to explore the air-borne transmission of viruses from a communication engineering perspective.
To this end, a human test person emitted fluorescent dyes during an induced cough.
The particles' locations were captured in \cite{schurwanz21} by a high-speed camera.
The propagation of information molecules in \cite{schurwanz21} relied on the impulsive force applied during the molecule release.

\subsubsection{Confined Environments}\label{air:bounded}

To study confined air-based \ac{SMC} systems, experiments are conducted within pipes and boxes, which constitute the boundaries.
The \acp{Tx} and \acp{Rx} are either attached via confined boxes to a connecting pipe \cite{qiu14,cole09,giannoukos17} or placed within the pipe \cite{shakya18, mcguiness19,ozmen18}. 
Within the pipe, possibly background flow with adjustable flow speed exists, similar to real world ventilation pipes, which is either controlled by a fan \cite{shakya18, mcguiness19, ozmen18} or by a mass flow controller \cite{giannoukos17}.
The motivations of these air-based \ac{SMC} experiments are variegated. 

\textbf{Spray-based transmission:} In \cite{qiu14}, the different path loss for \ac{SMC} in a confined environment as compared to \ac{EM}-based communication was studied. Hereby, the quality of the received signal was evaluated for different metallic pipe topologies, which were straight, single bend (L-shape), double bend (Z-shape), and reverse bend (U-shape). It was shown experimentally that, for some long ($>$ \SI{3.8}{\meter}) bent pipes, \ac{EM}-based communication is unreliable, while air-based \ac{SMC} using alcohol molecules is slow but feasible. In particular, the experiments revealed that the decay time of the received signal was dependent on the pipe shape and limited the data rate of the transmission in confined air-based \ac{SMC}.

\textbf{Air stream-based transmission with photo ionization detection:} In \cite{ozmen18, shakya18}, isopropyl alcohol molecules were employed as information molecules.
They were released into the pipe-shaped communication channel by a computer-controlled valve and sensed by photoionization detectors at the \ac{Rx} site.
A photoionization detector uses high-energy photons to break the molecules into positively charged ions. The ions produce a molecule concentration dependent current, which is measured by the detector.
In \cite{ozmen18}, a Karhunen-Loeve transform was performed based on pilot data, which captured the channel statistics.
By the transform, the stochastic process of the end-to-end information transmission was represented as an infinite linear combination of orthogonal functions denoted as motifs.
Based on these motifs, detection methods were developed.
The authors demonstrated in \cite{ozmen18} that relatively large data rates could be achieved, i.e., data rates of $40\,\mathrm{bps}$ at a \ac{BER} of $10^{-2}$ were shown to be possible.
In nature, low concentrations of high-frequency odor pulses can be detected due to high temporal resolution of the olfactory systems in insects \cite{szyszka14}. Motivated by this insight, high-frequency pulses of information molecules were used in \cite{shakya18} for data transmission. The authors showed that by matched filtering with the high frequent signal successful detection was possible even for settings with low transmit signal concentrations and background fnoise. In particular, the authors state that detection was possible nearly two orders of magnitude below the nominal detection limit of the sensor. 

\textbf{''Moth-on-a-chip'':} In \cite{cole09}, the biosynthesis of pheromones and their detection was investigated, whereas inspiration was drawn from pheromone signaling of the moth \textit{Spodoptera littoralis}. In particular, the \ac{Tx} synthesized different moth specific pheromone components, which were released via vaporization. The ligand binding at the olfactory receptors was implemented using a parallel array of human embryonic kidney cells, which carried the receptors. The binding of the ligands could be detected via surface acoustic wave devices.
Furthermore, a neuronal post-processing model, which mimics the signal processing in the antenna lobes, i.e., the olfactory brain area of insects, had been developed. The authors state that the experiment in \cite{cole09} aims to obtain a ''moth-on-a-chip'' to support pest management research in the long run.

\textbf{Air stream-based transmission with mass spectrometer detection:} In \cite{mcguiness19} and \cite{giannoukos17}, acetone molecules and odors (single molecules or mixtures), respectively, were released as \acp{VOC} from an evaporation chamber, transmitted over a pipe-shaped $\mathrm{N}_2$ gas stream channel, and detected by a membrane inlet mass spectrometer. The same authors analyzed the same system without a pipe in \cite{mcguiness18}.
In \cite{mcguiness19}, an analytical channel model was presented for the confined air-based \ac{SMC} systems.
Hereby, the channel boundaries were incorporated by the method of images.
The authors stated that their theoretically derived received signals showed general agreement with the experimentally obtained results.
Furthermore, a difference in the maximum signal amplitude was observed, which may have been caused by interactions between the membrane used in the detector and acetone.
In \cite{giannoukos17}, an end-to-end odor-based wireless transmission was proposed as ``odornet'', which allows for transmissions up to a distance of \SI{3}{\meter}.
The odornet emitter is able to control the vaporization and release of up to three individual volatile information molecules, which are mixed together in a mixing chamber.
Experiments were conducted for varying concentrations of the molecules, vaporization temperatures, the inner diameter and the length of the tube, and the flow rate of the background gas stream.

\textbf{Fluorescent spray-based:} In \cite{bhattacharjee20}, fluorescein molecules dissolved in water were shot into a tube by a spray nozzle and detected using a light-emitting diode and a camera.
The propagation of the information molecules in the testbed proposed in \cite{bhattacharjee20} relies mainly on the initial velocity of the particles.
This is different from the other testbeds reviewed in this section that employ background flow for the transport of the information molecules.

\subsection{Summary}

In this section, we reviewed different experimental works on air-based \ac{SMC}.
We classified the experiments according to the existence of boundaries in the environment and the applied release mechanisms and sensing principles, cf. Table~\ref{tab:air_based:release_mechanism}.
The reviewed works presented promising experiments towards future air-based \ac{SMC} applications, including mobile robots, natural pheromone-based \ac{SMC}, and \ac{SMC}  based on high-frequency, low-concentration pulses of information molecules.

%% file: conclusions_long-range.tex
In this part of our survey of experimental works on \ac{SMC}, we have reviewed long-range \ac{SMC} systems.

First, we have discussed several main types of prototypes for long-range in-body \ac{SMC}.
We have categorized them according to the combination of information molecules and physical \acp{Rx} they comprise.
In summary, when we compare the experimental works on long-range in-body \ac{SMC} reviewed in this paper with the works on short-range in-body \ac{SMC} reviewed in Part I of this survey, we observe that a more elaborate communication theoretic system design has often been provided for the former ones as compared to the latter ones.
The design of efficient detectors, for example, is more frequently considered in long-range systems.
Furthermore, several extensions were reported for different main experimental long-range in-body \ac{SMC} prototypes to develop them further in terms of the achievable data rate or the reliability of the transmission.

We could identify the following future research directions in this field:

\textbf{Tx and Rx design:} Although several different experimental works on long-range in-body \ac{SMC} exist, few have considered the design of practical \acp{Tx} and \acp{Rx}.
However, for the transition of \ac{SMC} towards a practical communication paradigm for in-body communication, such designs need to be developed.

\textbf{Realistic physical channels:} The experimental works on long-range in-body \ac{SMC} mimic the human cardio-vascular system with artificial tube-structures that oversimplify the real-world environment.
First, the material of the tubes is very different from the tissue that surrounds blood vessels.
Consequently, it is likely that the interaction of information molecules with the physical channel in a real-world application will be different from the interaction with the experimental environment, e.g., due to different friction coefficients at the respective channel boundaries.
Second, the topologies, i.e., junctions, etc., of the propagation channels considered in \ac{SMC} experiments are much simpler than the real-world channels.
It seems likely, however, that the topology of the channel has a significant impact on the signal propagation and that oversimplified models potentially underestimate the propagation loss in real-world channels.
In the future, sophisticated additive manufacturing techniques may provide more complex channel structures which are more similar to biological tissue.
Such models can help to improve the prediction of real-world channels in better fitting testbeds.

\textbf{\ac{SMC} in the blood stream with volatile and odorous substances and external detection:} Another direction for future research are volatile and odorous substances traveling the human body with the blood stream. Some of those substances serve as means of communication, e.g., about nutritional or health status, and are even used as diagnostic markers, e.g., in relation to the communication and detection of cancer. In this example, \ac{SMC} crosses both media, i.e., odorous and volatile signals pass from aqueous to gaseous phase, thereby transmitting their information.
Since invasive sensing devices are prone to the risk of infections, medical experts seek to exploit the aforementioned phenomenon to develop external non-invasive sensors, e.g., for the early detection of diseases.
\Ac{SMC} could provide a valuable tool for the design of such systems.

We have also reviewed experimental works on air-based \ac{SMC} in this paper.
The main combinations of release mechanisms and sensing principles for this type of \ac{SMC} systems were discussed and the reviewed experimental contributions were categorized accordingly.
Based on our survey, we have identified the following potential directions for future research in this area:

\textbf{Generalization of the conceptual framework for air-based \ac{SMC}:}
In future, this categorization might be based on general principles of \ac{SMC} rather than individual technological solutions.
Such principles include physico-chemical, biological and physiological principles, and categorization could be achieved by considering, e.g., temporal and spatial aspects for the signal transmission by the \ac{Tx}, and, e.g., specificity, sensitivity, and response behavior for the \ac{Rx}. 

\textbf{Natural olfaction-inspired transmission and sensing for air-based SMC:} The next generation of air-based \ac{SMC} systems could benefit from \ac{NMC} systems like olfaction and other natural detection systems of airborne molecular information. In these systems, multiple examples exist for different release mechanisms allowing for a temporally and spatially precise delivery of information, which ideally is robust under different environmental conditions. Analogously, different sensing principles have evolved, ranging from variously shaped detection organs to specific sampling behavior.
Air-water transitions also play an important role here. The mammalian olfactory epithelium, for instance, is an aqueous phase which serves pre-concentration and transformation purposes.

\textbf{Requirements engineering:} At the current stage of the research on air-based \ac{SMC} systems, it is very difficult to assess the quality and performance of a specific system design.
This is mainly due to the fact that the requirements imposed by specific target applications are not clear.
Carefully defined design goals and constraints, respectively, are, however, required as selectors for the further evolution of practical air-based \ac{SMC} prototypes.

\textbf{Optimization of mixed systems based on \ac{SMC} research:} We consider here as mixed systems those molecular communication systems that involve either a natural \ac{Tx} or a natural \ac{Rx}. Examples for such mixed systems are numerous, some have also been mentioned above as part of the motivation for studying air-based \ac{SMC} systems. Future research driven by the \ac{SMC} community, including theory-driven optimization of \ac{Tx} or \ac{Rx} designs, could help to further advance these mixed systems. For example, this could help to reliably monitor food quality before and after harvesting by optimizing the parameters relevant for the signal transmission between synthetic \acp{Rx} (e.g., used for measuring molecules indicating ripeness or spoilage) and the natural \acp{Tx} (e.g., fruits in the field or in the warehouse). Similarly, synthetic \acp{Tx} (e.g., pheromone traps) could be optimized in a way that they can be safely used and have maximum efficiency under different environmental conditions.

Despite the existing research challenges, we believe that future experiments together with theoretical modeling will reveal the operational principles of air-based \ac{SMC} that will eventually be the basis for the development of advanced technologies allowing for precise delivery and sensing of airborne chemical information in different surroundings.